# Multiscale dynamical network mechanisms underlying aging from birth to death

M. Zheng[1,+], Z. Cao[1,+], Y. Vorobyeva[2], P. Manrique[1], C. Song[1,3] and N.F. Johnson[1,3]

[1]Department of Physics, University of Miami, Coral Gables, FL 33146, USA
[2]Department of International Studies, University of Miami, Coral Gables, FL 33146, USA
[3]Complexity Initiative, University of Miami, Coral Gables, FL 33146, USA
+ These authors contributed equally to this work

**How self-organized networks develop, mature and degenerate is a key question for sociotechnical[1-4], cyber-physical[5-18] and biological systems[19-33] with potential applications from tackling violent extremism[34-41] through to neurological diseases[19,20]. So far, it has proved impossible to measure the continuous-time evolution of any *in vivo* organism network from birth to death[6,19,20,22]. Here we provide such a study which crosses all organizational and temporal scales, from individual components ($\sim O(1)$) through to the mesoscopic ($\sim O(10^3)$) and entire system scale ($\sim O(10^6)$). These continuous-time data reveal a lifespan driven by punctuated, real-time co-evolution of the structural and functional networks. Aging sees these structural and functional networks gradually diverge in terms of their small-worldness and eventually their connectivity. Dying emerges as an extended process associated with the formation of large but disjoint functional sub-networks together with an increasingly detached core. Our mathematical model quantifies the very different impacts that interventions will have on the overall lifetime, period of initial growth, peak of potency, and duration of old age, depending on when and how they are administered. In addition to their direct relevance to online extremism, our findings offer fresh insight into aging in any network system of comparable complexity for which extensive *in vivo* data is not yet available[6,7,19,20,30,31].**

It may be a long time before the dynamical evolution of any biological network is known at every instance of an organism's lifetime. Yet new insight is urgently needed for more incisive treatments against diseases such as Alzheimer's[19,20]. Likewise, in the field of counter-terrorism and societal security, understanding how online support of an extremist entity such as ISIS evolves over time could offer strategic benefits for knowing when and how to intervene[34-41]. Given that diverse complex systems have statistical similarities in their time-aggregated networks, and given the generic definition of an organism as a system of interdependent parts, a deeper understanding of the continuous-time dynamics in one specific system could yield insight into other networks of comparable complexity[3,6,7,8]. Motivated by this, we analyzed the body of pro-ISIS support that developed organically on VKontkate ([www.vk.com](www.vk.com)), and which made VKontakte a dominant social media site for ISIS recruitment, propaganda and financing[34,35] (see Supplementary Information SI). Our continuous-time study of its entire life cycle from initial growth (late 2014) until eventual death in late 2015[37,41], complements and extends existing landmark studies of dynamical networks such as the evolution of communications networks by Barabasi et al.[1], the annual garment industry decline by Uzzi et al.[10], aging in biological networks by Witten et al.[19,20], and neurological networks by Sporns and Bassett et al.[22-24]. Our study avoids the usual difficulties in defining network links and nodes that arise for spatiotemporally aggregated data, and the arbitrariness of manually imposed thresholds. Our analogies to the brain are meant to be descriptive, not statements of rigorous physical equivalence.

This entire lifetime of continuous-time network dynamics is shown in the online movie (SI) with snapshots in Fig. 1(a-c) during organism growth, maturity, and old-age approaching death. Each small node is a single user (akin to a neuron nucleus) who can generate links (akin to an axon) that can connect into a functional unit (i.e. online group as on Facebook, akin to a synapse: shaded larger circles in Fig. 1) in order to engage with other users (nodes) at arbitrary spatial distances (Fig. 1(d)). There are 203 functional units (online groups) and 97,679 nodes (users) appearing during the organism's lifetime. Akin to a synapse, each functional unit is self-organized and autonomous: it develops its own purpose (e.g. narrative concerning a recruitment campaign in a particular region, or operational information[34]) yielding not only redundancies but also substantial heterogeneity in terms of function. Like any living organism, the system has to evolve under continual endemic stress[37,39,41] – specifically, VKontakte moderators who sporadically shutdown functional units that promote pro-ISIS violence[37,39,41]. Each functional unit (and hence the function that it provides) can come into, and go out of, existence on very short or long timescales; can be small or large; and can change either very quickly or slowly over time. By analogy, this would suggest support for the conjecture that synapses in a brain are relatively unstable entities and hence unlikely as sites of long-term system memory[22]. The nodes by contrast are structural in that they provide the physical substance of the system, with new ones appearing mostly during the growth period and then largely remaining through death. Because of common nodes (users), the functional units develop interconnected modules or communities-of-communities (i.e. clusters of functional units) like a brain, allowing bottom-up transient coordination between functional units with rapid and efficient sharing of information, while also maintaining some functional specialization by creating boundaries that restrict the spread of information across the entire network[22].

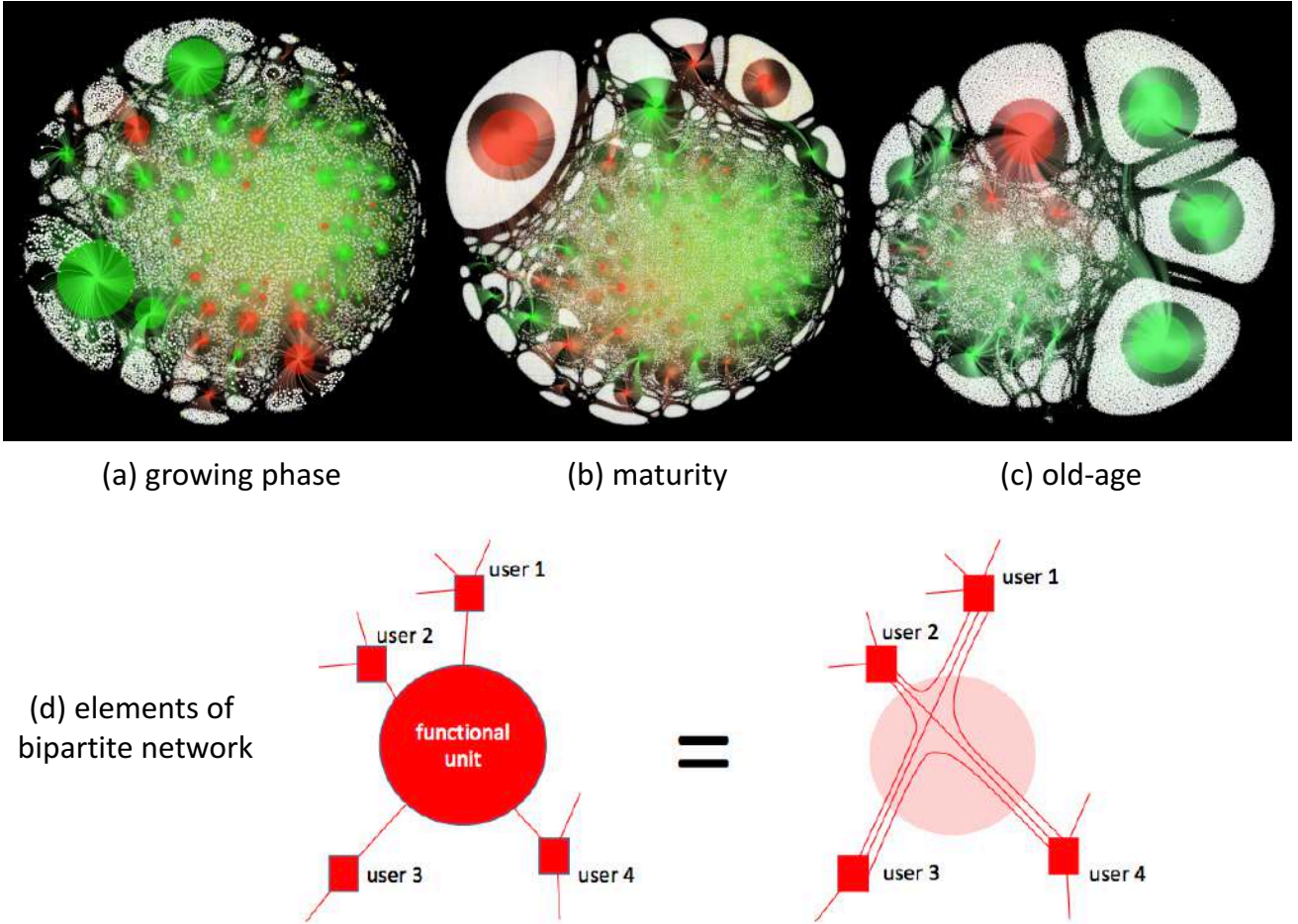

**Figure 1:** Organism network evolution. (a) Snapshots of the entire organism's bipartite network during (a) growth (example from day 25), (b) maturity (from day 125) and (c) old-age approaching death (from day 315). The smallest circles are individual users (nodes): yellow ones are users whose account eventually get banned for violating VKontakte's rules forbidding the promotion of pro-ISIS violence, while white ones are users whose account does not get banned. All bigger circles are functional units. Each user (i.e. smallest circles) in (a)-(c) can link into (i.e. follow) any number of functional units as shown schematically in (d). The functional units (i.e. larger circles) in (a)-(c) are shown as red if the corresponding online group gets banned for violating VKontakte's rules forbidding the promotion of pro-ISIS violence; and green if it does not get banned. See Ref. 34 and SI for more details and examples of these online groups. Overall, 203 functional units (online groups) and 97,679 nodes (users) appear during the organism lifetime.

Since the ForceAtlas2 algorithm used to generate Fig. 1(a-c) simulates a physical system in which nodes repel each other, functional units that have very few members in common and hence likely lack synchrony of function, will be shown as pushed apart in Fig. 1(a). During early growth (Fig. 1(a)) this rarely happens: instead, nodes are connected into multiple functional units, giving an overall synchrony but also affording less specialization in terms of distinctive function. The number of functional units grows fast and is almost proportional to the growth in number of member nodes, resulting in slow average growth of the functional units. The nodes show no strong preference for which functional unit to join, hence the functional units are close to each other. The organism also forms sub-networks — communities of communities. By maturity (Fig. 1(b)), a few larger distinct functional units have formed, suggesting some organism-specific functional capacities. However, the majority of functional units still share many members. Overall, this gives the system an advantageous blend of specialization with overall synchrony. By old-age and approaching death (Fig. 1(c)), several giant functional units dominate, but their position on the boundary indicates that they share very few common nodes. Moreover, the core has become visibly sparser and more detached, as confirmed quantitatively by the correlation matrices for functional units (Fig. 2), as well as the sharp increase in the fraction of nodes involved in just one functional unit (see SI). The process of advanced aging toward death (Figs. 2(c) and (d)) reduces the overall resilience to possible damage, since the organism lacks the synchrony (i.e. common nodes) between functional units to substitute and coordinate the possible loss of a given unit. By contrast during early growth (Fig. 1(a) and 2(a)), the functional clusters that emerged had common members and hence could have likely substituted function – albeit at the expense of efficiency because of the high overlap, and an increased susceptibility to confusion of function[22].

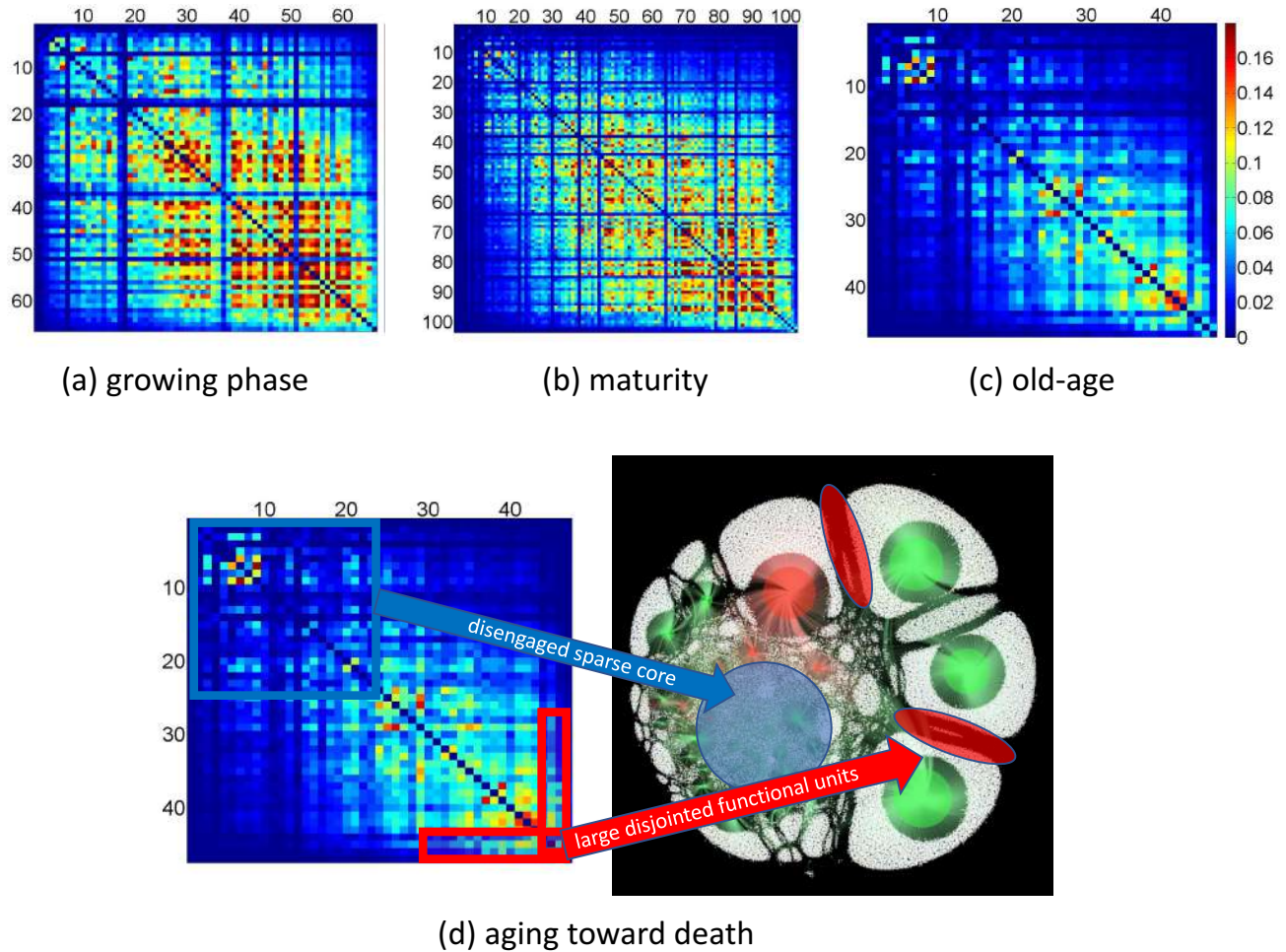

(a) growing phase    (b) maturity    (c) old-age

(d) aging toward death

**Figure 2:** Aging and dying. (a)-(c) Correlation matrices correspond to the snapshots (a)-(c) in Fig. 1. They show the overlap between different functional units at time $t$, which is given by the Jaccard index, i.e. entry $(i,j)$ at time $t$ is given by the number of common users in functional units $i$ and $j$ divided by the total number of unique members in both functional units $i$ and $j$, all measured at time $t$. The dimension of the matrix $i,j = 1,2,3,...n(t)$ is the number of functional units at that time $t$, and they are ordered by increasing size. (d) Process of dying involves formation of a disengaged but sparse core, together with large but disjointed functional units.

Figure 3 provides continuous-time network measures for the organism after projection at each timestep onto the users, giving the structural network (Fig. 3(a)), and onto the functional units, giving the functional network (Fig. 3(b)). The emergent periods of growth, maturity, and degeneration are characterized by short-lived trends punctuated by frequent shifts in each measure. The time-dependencies highlight the remarkable tendency for rewiring and remodeling at the scale of individual nodes (users) and functional units (groups), which crosses scales to become a punctuated co-evolution of the functional and structural networks. Many of the gross structural and functional network features are consistent with broad properties conjectured for brain networks[22]: e.g. high clustering and short average path lengths; broad degree distribution with small subsets of highly connected nodes (hubs); rich-club organization; and communities of communities (i.e. modules of functional units). However, Fig. 3(a-c) also reveals a new form of dynamical, intra-organism competition underlying the passage through life towards death: The functional network undergoes a decreasing overall trend in clustering coefficient (i.e. modularity) accompanied by a general increase in its average path length (Fig. 3(b)), indicating an overall loss of small-world behavior which makes it harder for distant functional units to know about each other. By contrast, the structural network (Fig. 3(a)) undergoes an increase in clustering coefficient and a decrease in average path length, which means that the small-world behavior increases -- hence any attack on one hub node can more easily transfer to another. In addition, the average degree in the functional network decreases markedly into old-age while that of the structural network appears to saturate (Fig. 3(c)). Irrespective of whether the organism's growth and subsequent decline toward death are measured in terms of the number of underlying follows as in Fig. 3(d) (i.e. number of links created by the nodes), or the number of functional units (online groups), or the total number of users, or the number of new postings, or the number of group-joining events, all these measures rise and fall over the lifetime and all die out by the end of 2015. But they do so in different ways and with regime changes at different times, hence generating the complex dynamical interdependencies in Figs. 1 and 2.

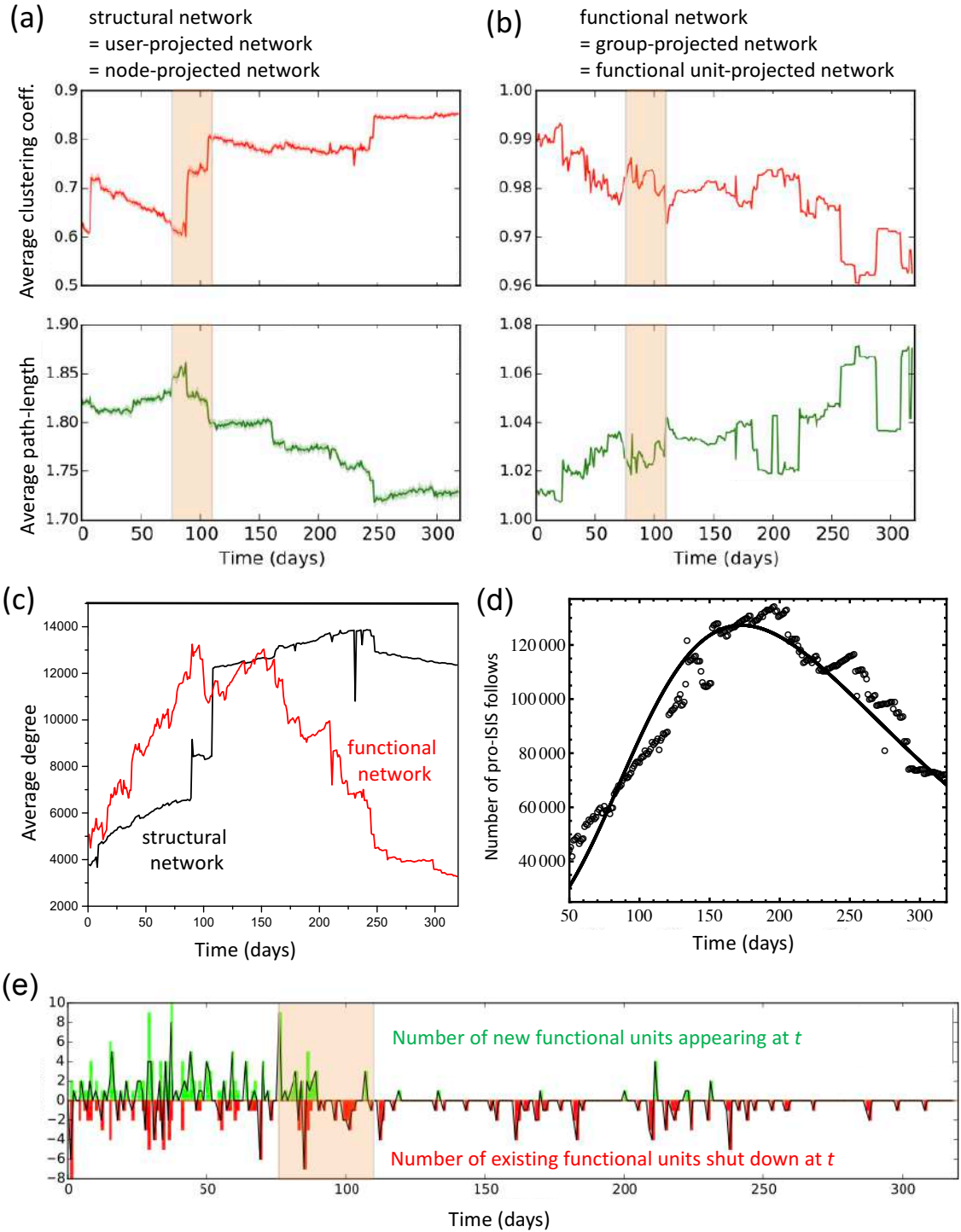

**Figure 3:** Continuous-time network properties. Temporal variation of the network properties from Fig. 1(a) after projecting onto (a) the nodes (i.e. users) giving the structural network, and (b) the functional units (i.e. online groups) giving the functional network. While (b) is for the entire network, the huge number of users (nodes) led us in (a) to sample 5000 users 10 times, hence the 1-sigma error band shown. We checked that increasing this sample size did not change our results or conclusions. The regime of trauma $76 < t < 110$ is shown as shaded orange. (c) Temporal variation of the average degree for the structural and functional networks. (d) Temporal variation of the total number of pro-ISIS follows. Points show empirical data. Line shows result from our mathematical model involving generation of pro-ISIS follows as a result of transience through the set of pro-ISIS groups. Our model is generalized from Ref. 42 and uses realistic parameters. (e) Temporal variation in the number of new functional units appearing (top: green) and disappearing (bottom: red) at time $t$. Black line is the net change (i.e. difference).

'Dying' hence emerges as a process that is both asynchronous and extended: different network features either die out or saturate at different rates and times with some processes starting early in the lifespan (Fig. 3), and large but disjoint functional units eventually develop (Fig. 1-2). While there is some sense of structural homeostasis in that the average degree and the number of nodes in the structural network roughly saturate in old age (Fig. 3(a)), the functional network shows no such saturation. As death approaches, the organism appears to fight back (see Fig. 3(e) $t \approx 220$) with a burst of new functional units, and an associated burstiness in the functional network average path-length and clustering (Fig. 3(b)) as if the organism were trying -- ultimately unsuccessfully -- to compensate for the loss of other functional units by increasing its small-world functional behavior. Further evidence that that the ability to adapt becomes weaker over time, i.e. reduction in plasticity, is shown by the progressive reduction in the rate of change of the structural network clustering coefficient following each of the four major jumps in Fig. 3(a). The organism's shift from growth to degeneration toward death, is associated (Fig. 3(e)) with a fall in the appearance of replacement functional units, as opposed to a rise in the destruction of existing functional units. The organism compensates this overall reduction by a general increase in the average size of functional units (Fig. 1(c)). However, this attempt at improving system-wide coordination is hampered by the increasing functional network path-length and its decreasing clustering coefficient.

Surprising bottom-up coping strategies emerge in response to both the endemic stress of moderator shutdowns (Fig. 3(e)) and the major one-off trauma at timestep $t \approx 76$. For example, new functional units of considerable size occasionally appear from out of nowhere. While it is possible that the member nodes quickly self-organize to produce them, deeper investigation of the data shows that they can also originate from latent functional units which were previously unengaged in terms of pro-ISIS support, but which have their connections already in place and hence can quickly spring into action. This differs from, and is more common than, the creation of multiple user accounts as on Twitter, and raises the question whether such 'latent synapses' lie hidden – possibly untapped -- in natural systems such as the brain. The response to the trauma is equally novel: Though practically every day saw real-world actions against ISIS, the wounding of ISIS leader Abu Bakr al-Baghdadi in a Coalition airstrike on timestep $t \approx 76$ (March 18, 2015) was the most traumatic in that it was the only one that directly impacted ISIS' leadership. Rumors immediately circulated among some functional units that the top ISIS leaders were meeting to discuss who would replace him if he died, suggesting that his injuries were serious. However, none of this become public knowledge in the media until $t \approx 110$ when both the $t \approx 76$ attack and the unexpected seriousness of his wounds were reported. This period of permeating rumors ($76 < t < 110$) coincides with an increase in the frequency of punctuated internal shifts in the structural network clustering (Fig. 3(a) orange shaded region) which allowed the system to globally re-structure itself. Moreover, the organism's immediate response at $t = 76$ included generating new functional units using this 'latent synapse' effect.

External interventions will have very different outcomes depending on when – and how – they are administered. To quantify this, we developed a coupled differential equation model following Ref. 42 which yields good agreement with the data for the overall lifespan (Fig. 3(d)). It features follows joining the online space dynamically in time, and then becoming infected (i.e. the link joins into a pro-ISIS functional unit). The main features of Fig. 3(d) are the time-to-peak $T_m$, the peak in potency (i.e. peak height) $H$, and the lifetime $T$. Figure 4 shows how these are altered by changing the intervention time $t_I$ and size, where each intervention involves a fraction of follows being randomly removed at time $t_I$. An early intervention (e.g. $t_I = 80$) *prolongs* the lifetime $T$; a slightly later one ($t_I = 120$) leaves it statistically unchanged; but one that is even later (e.g. $t_I = 200$) *reduces* the lifetime $T$ (see SI for specific numbers). An early intervention (e.g. $t_I = 80$) increases the time-to-peak (i.e. duration of the growth phase) $T_m$ while a slightly later one (e.g. $t_I = 120$) decreases it. Finally, both an early intervention (e.g. $t_I = 80$) and a slightly later one (e.g. $t_I = 120$) decrease the peak height (i.e. potency) $H$, with the former having more impact.

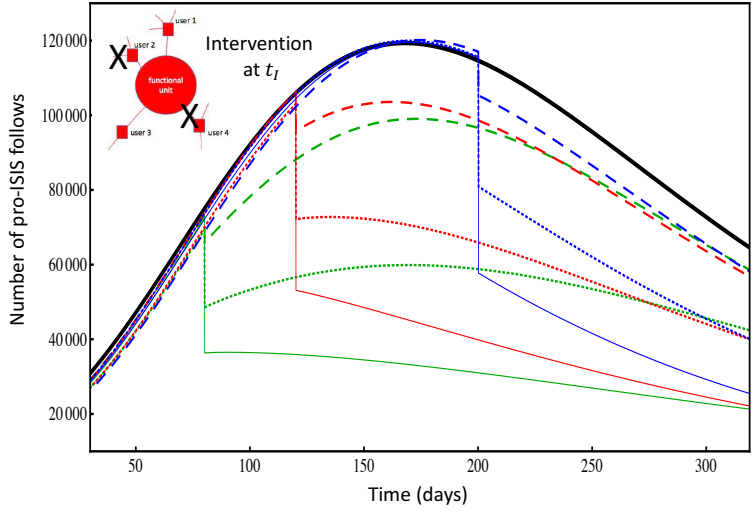

**Figure 4:** Impact of interventions. The impact of one-off intervention at time $t_I$, as described in the text and inset, is quantified by our mathematical model used in Fig. 3(d). The predicted impact of each separate intervention is indicated by its own jump and subsequent line modification. Each curve is an average over 150 simulations of the system, with intervention (colored curves) and without intervention (black curve from Fig. 3(d)). Colors represent the moment of intervention $t_I$ in units of days (vertical gray lines): green $t_I = 80$, red $t_I = 120$, blue $t_I = 200$. Line types represent the size of the intervention, i.e. percentage of follows that are randomly removed at time $t_I$: dashed line 10%, dotted line 30%, solid line 50%. The lifetime $T$ can be taken as the time until the number of follows falls to some small fraction of the total (e.g. 1%), though our conclusions are all robust to different choices of this fraction.

We also calculated analytically the time $T_e$ that it will take a specific 'treatment' (an external opposition network having a similar structure and dynamics but finite resources, i.e. fixed number of nodes $N_{ext}$) to eliminate *all* $N_{org}$ organism nodes (pro-ISIS users) starting from its maximum value, such that no organism node lies buried or dormant. $T_e$ is given by a generalization of Ref. 43 in terms of derivatives of the logarithm of a Gamma function $\Psi(z) = d\ln\Gamma(z)/dz$:

$$T_e = \frac{N_{ext}-N_{org}}{2c}\left[\frac{4N_{org}}{N_{ext}-N_{org}} - \Psi\left(\frac{N_{ext}-N_{org}+c}{c}\right) - \Psi\left(\frac{N_{ext}+c}{c}\right) + \Psi\left(\frac{N_{org}+c}{c}\right) + \gamma\right] \quad (1)$$

where $\gamma$ is the Euler-Mascheroni constant and $c$ is the average number of organism nodes destroyed per encounter. Because this takes into account both the time needed to find and destroy all organism nodes, including isolated ones, $T_e$ has a maximum when the fraction of organism nodes is small compared to the attackers (i.e. when $x =$ $(N_\text{ext} - N_\text{org})/(N_\text{ext} + N_\text{org})$ is near to 1). This means that a very small online organism will take far longer to completely eliminate than might otherwise be expected based purely on having a majority of opponents ($N_\text{ext} > N_\text{org}$), meaning that small extremist entities are well-suited to online survival. However, an important policy implication from Eq. (1) is that the value of $x$ at which the maximum in $T_e$ arises, varies approximately as $x_\text{max} \approx \frac{1}{6}\left(\sqrt{3\log\frac{N_\text{ext}+N_\text{org}}{8c}-5}+2\right)$ for small $c$ which means that a suitable value of $N_\text{ext}$ can always be chosen in order to avoid the longest elimination time.

**Acknowledgments:** NFJ is funded by National Science Foundation (NSF) grant CNS 1522693 and Air Force (AFOSR) grant FA9550-16-1-0247. The views and conclusions contained herein are solely those of the authors and do not represent official policies or endorsements by any of the entities named in this paper.